\def\Granadadep{Departamento de F\'\i sica Te\'orica y del Cosmos,
Facultad
de Ciencias, Universidad de Granada, Campus de Fuentenueva, Granada
18002, Spain}

\def\Granadainst{Instituto de F\'\i sica Te\'orica y Computacional
Carlos I, Facultad
de Ciencias, Universidad de Granada, Campus de Fuentenueva, Granada
18002, Spain}

\def\Valencia{IFIC, Centro Mixto Universidad de Valencia-CSIC,
Burjasot
              46100-Valencia,Spain.}

\def\nn{\nonumber}
\def\ni{\noindent}
\def\te{\tilde{\eta}}
\def\tz{\tilde{\zeta}}
\def\tc{\tilde{C}}
\def\tcc{\tilde{C^*}}
\def\tw{\theta_W}

\documentstyle[12pt]{article}

\def\Comision{Work partially supported by the DGICYT.}

\title{ GROUP-THEORETICAL DETERMINATION OF THE MIXING ANGLE IN THE
        ELECTROWEAK GAUGE GROUP\thanks{\Comision} }

\author{ V. Aldaya$^{1,2}$  \and M. Calixto$^{1}$  \and J.
Guerrero$^{1,3}$}

\date{}

\begin{document}

\setcounter{page}{0}

\footnotetext[1]{\Granadainst} \footnotetext[2]{\Valencia}
\footnotetext[3]{\Granadadep}

\maketitle

{\bf Abstract. }{\small The assumption that the Weinberg rotation
between
the gauge fields associated with the third component of the ``weak
isospin"
($T_3$) and the hypercharge ($Y$)  proceeds in a natural way from a
global
homomorphism of the $SU(2)\otimes U(1)$ gauge group in some locally
isomorphic
group (which proves to be $U(2)$), imposes strong restrictions so as
to fix the single value
$\sin^2\theta_W=1/2$. This result can  be thought of only as being an
asymptotic
limit corresponding to an earlier stage of the Universe. It also
lends support
to the idea that $e^2/g^2$ and $1-M_W^2/M_Z^2$ are  in principle
unrelated quantities.}

\vskip 1cm

There are two basic ingredients in the constitution of a model to
describe the
unified electroweak interactions, the Weinberg-Salam-Glashow Standard
Model,
which  deserve further study and which lessen the (mathematical)
beauty of the theory as a whole. One is the way in which the
W-Z-bosons acquire
mass, the Higgs mechanism, and the other is the rotation between the
gauge
fields associated with the third component of weak isospin
($A^3_\mu$) and the
hypercharge ($A^4_\mu$), intended to define the proper
electromagnetic field,
without any (apparent) connection to the ``weak" Gell'man-Nishijima
relation

\begin{equation}
Q=T_3 +\frac{1}{2}Y \, ,
\end{equation}

\noindent meant to define a proper electric charge in the
Lie algebra. We shall focus on the latter question.

In this paper we wish to explore the restrictions that appear on the
mixing
angle $\theta_W$ as a consequence of the natural consistence
requirement
that the rotation in the gauge fields

\begin{eqnarray}
Z^0_\mu &=& \cos\theta_W A^3_\mu - \sin\theta_W A^4_\mu \\
A_\mu &=& \sin\theta_W A^3_\mu + \cos\theta_W A^4_\mu \nn
\label{Wrotation}
\end{eqnarray}

\noindent comes from an exponentiable (linear) transformation on the
Lie algebra
of $SU(2)\otimes U(1)$. Since the gauge group is not simply
connected, it is not
true that any automorphism of the Lie algebra can be realized as the
derivative
of a global group homomorphism or, in other words, a differentiable
mapping between
two locally isomorphic groups providing a given automorphism of the
(commom) Lie
algebra, can in general destroy the global group law.


 To analyse the set of global homomorphism from $SU(2)\otimes U(1)$
to a
locally isomorphic group we can proceed in two different ways: either
we study
the set of discrete normal subgroups of $SU(2)\otimes U(1)$, which
are the possible
kernels of those homomorphisms, or we write the explicit group law of
$SU(2)\otimes U(1)$,
perform an arbitrary homomorphism and analyze the conditions under
which the group law
is not destroyed. We shall follow the second approach although some
comments on the
first one will be added at the end.

Let us parametrize the group $SU(2)$ in a co-ordinate system adapted
to the
Hopf fibration $SU(2)\rightarrow S^2$, the sphere $S^2$ being
parametrized by
stereographic projection. The  $SU(2)\otimes U(1)$ group law in the
local chart at the
identity, which nevertheless keeps the global character of the toral
subgroup, is:

\begin{eqnarray}
\eta''
&=&\frac{z_1''}{|z_1''|}=\frac{\eta'\eta-\eta'\eta^*C'C^*}{\sqrt{(1-
{\eta^*}^2C'C^*)(1-\eta^2C{C^*}')}}  \nn \\
C''&=&\frac{z_2''}{z_1''}=\frac{C\eta^2+C'}{\eta^2-C'C^*}  \\
{C^*}''&=&\frac{{z_2''}^*}{{z_1''}^{*}}=
   \frac{C^*\eta^{-2}+{C^*}'}{\eta^{-2}-{C^*}'C}  \nn \\
\zeta''&=&\zeta'\zeta \nn
\end{eqnarray}

\ni where $\eta \in U(1)\subset SU(2), \zeta \in U(1),
C\in$ {\bf C} and $z_1, z_2$ characterize a $SU(2)$ matrix $\left(
\begin{array}{cc}
z_1&z_2\\-{z_2}^*&{z_1}^*\end{array}\right)$. The commutation
relations between
the (right) generators $T_+ \equiv X_{C^*},\,T_-\equiv X_C, T_3
\equiv X_\eta$ and
$Y \equiv X_\zeta$ are:

\begin{eqnarray}
\left[T_3,T_\pm \right] &=& \pm 2 T_\pm \nn \\
\left[T_+,T_- \right] &=&T_3 \\
\left[Y,\hbox{all} \right] &=& 0 \nn
\end{eqnarray}

\vskip .25cm
We shall consider transformations induced by an homomorphism of the
torus:

\begin{eqnarray}
\tilde{\eta}&=&\eta^p\zeta^{p'} \nn \\
\tilde{\zeta}&=&\eta^q\zeta^{q'} \label{toro} \\
\tilde{C}&=& C \, , \, \, \tilde{C^*}=C^*, \nn
\end{eqnarray}

\ni where the parameters $p, p', q, q'$ have to be integers for the
univalueness
requirement.

After we apply this transformation the group law becomes:

\begin{eqnarray}
\te''&=&\left (\frac{\te'{}^\frac{1}{p}\te^\frac{1}{p}-
  \te'{}^\frac{1}{p}\te^{-\frac{d+2qp'}{dp}}\tz^\frac{2p'}{d} \tcc
\tc'}
  {\sqrt{(1-\te^{-\frac{2q'}{d}}\tz^\frac{2p'}{d} \tc' \tcc)
  (1-\te^\frac{2q'}{d}\tz^{-\frac{2p'}{d}} \tc \tcc')}}\right ) ^p
\nn  \\
\tc''&=&\frac{\tc \te^\frac{2q'}{d}\tz^{-\frac{2p'}{d}}+\tc'}
        {\te^\frac{2q'}{d} \tz^{-\frac{2p'}{d}}- \tc' \tcc} \\
\tcc''&=&\frac{\tcc \te^{-\frac{2q'}{d}} \tz^\frac{2p'}{d}+\tcc'}
         {\te^{-\frac{2q'}{d}} \tz^\frac{2p'}{d}-\tcc' \tc}
 \nn  \\
\tz''&=&\tz'\tz(\te''\te'{}^{-1}\te{}^{-1})^\frac{q}{p}  \, \, , \nn
\label{lawtilde}
\end{eqnarray}

\ni where $d$ is the determinant of the matrix  $\left(
\begin{array}{cc}
p&p'\\q&q'\end{array}\right)$,  and this group law is well-behaved if

\begin{equation}
\frac{2p'}{d}=m, \, \, \frac{2q'}{d}=n, \, \, \frac{q}{p}=k, \, \, m,
n, k \in Z
\label{diofantine}
\end{equation}

\ni which, in particular, imply $p=\pm 1, \pm 2$. This particular
result simply
states the well-known fact that the only invariant subgroups of
$SU(2)$ itself
are $I$ (the identity) and $Z_2$, respectively.

The commutation relations between the new generators (with a
definition analogous
to that given above),

\begin{eqnarray}
\left[\tilde{T}_3,\tilde{T}_\pm \right] &=& \pm \frac{2q'}{d}
\tilde{T}_\pm \nn \\
\left[\tilde{Y},\tilde{T}_\pm \right] &=& \pm \frac{-2p'}{d}
\tilde{T}_\pm  \\
\left[\tilde{T}_+,\tilde{T}_- \right] &=& p \tilde{T}_3 + q \tilde{Y}
\nn \,,
\end{eqnarray}

\ni can be obtained directly from (\ref{lawtilde}) or by applying
the tangent mapping to (\ref{toro}) to the old ones. This
transformation gives:

\begin{eqnarray}
\tilde{T}_3&=&\frac{q'}{d}T_3-\frac{q}{d}Y  \nn \\
\tilde{Y}&=&\frac{-p'}{d}T_3+\frac{p}{d}Y \label{Gell}
\end{eqnarray}

\ni and provides a generalized Gell'Mann-Nishijima relation and its
counterpart,
which now appear quantized.

Let us now examine the transformation induced by (\ref{toro}) in the
($3^{rd}-4^{th}$
internal components of the) gauge fields. It is given by:

\begin{equation}
\left( \begin{array}{c} \tilde{A}^3_\mu \\ \tilde{A}^4_\mu
\end{array} \right)
= \left( \begin{array}{cc}
\frac{1}{\tilde{r}}&0\\0&\frac{1}{\tilde{r}'}\end{array}
\right) \left( \begin{array}{cc} p&p'\\ q&q'\end{array}
\right) \left( \begin{array}{cc} r&0\\0&r'\end{array}\right)
\left( \begin{array}{c} A^3_\mu \\ A^4_\mu \end{array} \right)
\label{rotacion}
\end{equation}

\ni where $r, r'$ are the original coupling constants associated with
isospin and hypercharge respectively, and $\tilde{r}, \tilde{r}'$ are
the final
ones. In fact, the covariant derivative
$D_\mu=\partial_\mu-ig^k_iT_kA^i_\mu$, where
$i,k$ run over 1,2,3,4 ($T_4\equiv Y$) goes to
$\tilde{D}_\mu=\partial_\mu-i\tilde{g}^k_i\tilde{T}_k\tilde{A}^i_\mu=
D_\mu$. Therefore,

\begin{equation}
\tilde{A}^l_\mu=(\tilde{g}^{-1}){}^l_j\,a^j_k\,g^k_i\,A^i_\mu
\end{equation}

\ni where $a^j_k$ is the transformation matrix changing co-ordinates
in the Lie algebra,
which contains the central matrix in (\ref{rotacion}) as a box, and
${\bf g}=\hbox{diag}(r,r,r,r')$ and
${\bf
\tilde{g}}=\hbox{diag}(\tilde{r},\tilde{r},\tilde{r},\tilde{r}')$ are
the
initial and final (bare) coupling constants matrices.

We now impose the requirement that the complete transformation
(\ref{rotacion}), rather than the
central matrix in it, be the Weinberg rotation (\ref{Wrotation})
($Z^0_\mu\equiv \tilde{A}^3_\mu, A_\mu\equiv\tilde{A}^4_\mu$). This
results in

\begin{equation}
\frac{\tilde{r}^2}{\tilde{r}'{}^2}=-\frac{pp'}{qq'}\,,\:\:
\frac{r^2}{r'{}^2}=-\frac{q'p'}{qp}\,,\:\:
\tan^2\tw=-\frac{qp'}{pq'}\,,\:\:
\tilde{r}=\frac{p}{\cos\tw}\;r \, ,\label{tangente}
\end{equation}

\ni which contain a further restriction: the product of the four
integers
$pp'qq' < 0$, a condition afterwards necessary to have a
(non-trivial)
rotation. If the transformation (\ref{toro}) is  an automorphism of
the torus ($d=\pm 1$), then the only possible rotations between the
gauge fields
are the trivial ones ($\tan^2\tw=0,\infty$), so that the final group
has to be
the quotient of $SU(2)\otimes U(1)$ by a non-trivial normal
(discrete) subgroup.
 Adding (\ref{tangente}) to (\ref{diofantine}) we arrive at the final
result:

\begin{equation}
\left\{ p=\pm 1 \: \: \hbox{and} \: \: (p'=-kq', k=\pm 1)\right\}
\Rightarrow
             \left\{ \tan^2\tw = 1 \, ,\, d=\pm 2 q' \right\}
\end{equation}

\ni For these values of $p,p',q,q'$ the kernel of the homomorphism
(see the
transformation (\ref{toro})) is the normal subgroup

\begin{equation}
H_d \equiv
\left\{(C,C^*,\eta;\zeta)=
(0,0,1;e^{i\frac{2s}{d}\,2\pi}),(0,0,-1;e^{i\frac{2s+1}{d}\,2\pi})
/ s=0,1,...,\frac{|d|}{2}-1\right\}
\end{equation}

\ni which is isomorphic, as a group, to $Z_{|d|}$. All these
homomorphisms lead to the same value for $\tan^2\tw (=1)$ and indeed,
all
can be written as:

\begin{equation}
\left( \begin{array}{cc}p&p'\\q&q'\end{array} \right) =
\left( \begin{array}{cc}\pm 1&-kq'\\ \pm k&q'\end{array} \right) =
\left( \begin{array}{cc}1&-k\\k&1\end{array} \right)
\left( \begin{array}{cc}\pm 1&0\\0&q'\end{array} \right)
\end{equation}

\ni where the second factor has determinant $\pm q'$, and represents
a
transformation from $SU(2)\otimes U(1)$ to $SU(2)\otimes
(U(1)/Z_{|q'|})$, and
the first one has determinant $2$ and  would take $SU(2)\otimes U(1)$
to $(SU(2)\otimes U(1))/H_2 \approx U(2)$ by itself. The second
factor
affects the quotient between the original coupling constants (not the
final one), as can be seen
in (\ref{tangente}), and the generalized Gell'Man-Nishijima relation
(\ref{Gell}). Among
the possible values for $q'$ only $q'= \pm 1$  provides us with a
proper
electric charge; the choice of the signs of $p,q,p'$ is a matter of
convention and will define either $\tilde{T}_3$ or $\tilde{Y}$ as
$\pm$ the electric
charge $Q$. The corresponding homomorphism has Kernel
$H_2=\left\{(0,0,1;1),
(0,0,-1;-1)\right\}$  and $U(2)$ as the image group (true gauge
group)
\cite{Isham,Lor,LaChapelle}.

With the usual choice of multiplets in the Lagrangian of the Standard
Model (see e.g. \cite{Ryder}) $T_3$ and $Y$ have the expressions

\begin{equation}
T_3=\left(\begin{array}{ccc} 1&0&0\\0&-1&0\\0&0&0
\end{array}\right)\:,\:
Y=\left(\begin{array}{ccc} 1&0&0\\0&1&0\\0&0&2 \end{array}\right)
\end{equation}

\ni which agrees with the usual expressions if the $U(1)$ subgroups
are
trivially reparametrized by $\alpha= -2i\ln\eta, \beta=i\ln\zeta$
($T_{3}\rightarrow \frac{1}{2}T_{3}, Y\rightarrow -Y$). The
particular
choice of signs $p=p'=q'=-q=-1$ yields:

\begin{equation}
 Q=\tilde{Y}=\left(\begin{array}{ccc} 0&0&0\\0&-1&0\\0&0&-1
\end{array}\right)\:,\:
\tilde{T}_3=\left(\begin{array}{ccc} -1&0&0\\0&0&0\\0&0&-1
\end{array}\right)
\end{equation}

The first surprising result is the fact that only one value of
$\tan^2\tw$
is allowed, which means only one coupling constant (the electric
charge, essentially, i.e.
$e\equiv \tilde{r}' = \sqrt{2}r\equiv g/\sqrt{2}$),
even though the gauge group ($U(2)$) is not a simple group. According
to general settings
\cite{Itzykson}, however, the theory must contain a coupling constant
for each simple
or abelian term in the Lie algebra decomposition. An immediate
conclusion is that
the assignment of constants should be done according to factors in
the direct product
decomposition of the group, rather than the algebra.

The second result is the particular structure of the neutral weak
current
derived from the expression of $\tilde{T}_3$ above, according to
which the
gauge field $Z^0$ interacts with the (left-handed) neutrino and the
right-handed electron only; i.e. the neutral weak current is pure V-A
for
the neutrino and pure V+A for the electron.

 Last, but not least, is the striking value of
$\sin^2\tw=\frac{1}{2}$ ($\tan^2\tw=1$), far from the experimental
value
$\approx 0.23$ \cite{tabla}. In the light of this result, only the
hope remains
that our theoretical value of $\tw$ really corresponds to that state
of the
Universe in which the electroweak interaction was not yet
spontaneously broken, and that the process of spontaneous symmetry
breaking,
not fully understood (at least from a pure group-theoretical point of
view)
could relax the strong conditions (\ref{diofantine}). For instance,
breaking down the $SU(2)$ group law (\ref{lawtilde}) and preserving
the
$U_{em}(1)$, leads to $\tan^2\tw=-\frac{q}{q'}$, allowing any
rational value.

In any case, the discrepancy between our theoretical value for
$sin^2\tw$ as
given by a ratio of coupling constants,
$\frac{e^2}{g^2}=\frac{\tilde{r}'{}^2}{(2r)^2}$, and the experimental
one
obtained through the expression $1-\frac{M_W^2}{M_Z^2}$, lends
support to the
idea that in principle both quantities are not  related
\cite{Veltman}.

\vskip 0.5 cm

\ni {\large {\bf Acknowledgements}} V.A. is grateful to M. Asorey, J.
Julve and
A. Tiemblo, and all of us to J.M. Cerver\'o, J. Navarro-Salas, M.
Navarro and
 A. Romero for valuable
discussions.


\newpage

\begin{thebibliography}{9}

\bibitem{Weinberg} S. Weinberg, Phys. Rev. {\bf 19}, 1264 (1967)



\bibitem{Isham}  C.J. Isham, J. Phys. A {\bf 14}, 2943 (1981)

\bibitem{Lor} L. O'Raifeartaigh, {\it Group structure of gauge
theories},
              Cambridge University Press (1986)

\bibitem{LaChapelle} J. LaChapelle, J. Math. Phys. {\bf 35}, 2186
(1994)

\bibitem{Ryder} L.H. Ryder, {\it Quantum Field Theory}, Cambridge
University
                Press (1985)

\bibitem{Itzykson} C. Itzykson and J.B. Zuber, {\it Quantum Field
Theory},
                MacGraw-Hill International Edition (1985)

\bibitem{tabla} M. Aguilar-Benitez et al., Phys. Rev. {\bf D45}, Part
2 (Review of Particle Properties) (1992)


\bibitem{Veltman} G. Passarino and M. Veltman, Phys. Let. {\bf B237},
537 (1990)

\end{thebibliography}
\end{document}